# Enhancing crystal structure prediction by combining computational and experimental data via graph networks

Chenglong Qin [a, 1], Jinde Liu [a, 1], Shiyin Ma [a], Jiguang Du [b], Gang Jiang [a], Liang Zhao [a, *]

[a] Institute of Atomic and Molecular Physics, Sichuan University, Chengdu 610065, China
[b] College of Physics, Sichuan University, Chengdu 610064, China

**Abstract**

Crystal structure prediction (CSP) stands as a powerful tool in materials science, driving the discovery and design of innovative materials. However, existing CSP methods heavily rely on formation enthalpies derived from density functional theory (DFT) calculations, often overlooking differences between DFT and experimental values. Moreover, material synthesis is intricately influenced by factors such as kinetics and experimental conditions. To overcome these limitations, a novel collaborative approach was proposed for CSP that combines DFT with experimental data, utilizing advanced deep learning models and optimization algorithms. We illustrate the capability to predict formation enthalpies that closely align with actual experimental observations through the transfer learning on experimental data. By incorporating experimental synthesizable information of crystals, our model is capable of reverse engineering crystal structures that can be synthesized in experiments. Applying the model to 17 representative compounds, the results indicate that the model can accurately identify experimentally synthesized structures with high precision. Moreover, the obtained formation enthalpies and lattice constants closely align with experimental values, underscoring the model's effectiveness. The synergistic approach between theoretical and experimental data bridges the longstanding disparities between theoretical predictions and experimental results, thereby alleviating the demand for extensive and costly experimental trials.

**Introduction**

The goal articulated by the Materials Genome Initiative (MGI) in 2011 is ambitious: "Discover, develop, manufacture, and deploy advanced materials at least twice as fast as is currently feasible today, at a fraction of the cost" [1]. Traditionally, the discovery of new materials relied on serendipity or trial-and-error methods, both demanding and labor-intensive processes. However, over the past decade, the advent of crystal structure prediction (CSP) methods based on computational means has revolutionized the systematic discovery of materials[2, 3, 4, 5]. The integration of first-principles computations with optimization algorithms, such as evolutionary algorithms, meta-dynamics, or particle swarm optimization, has led to significant advancements of diverse applications in CSP[6, 7, 8, 9, 10]. Presently, CSP is experiencing a surge in interest due to significant breakthroughs across diverse fields such as drug design, high-pressure chemistry, computational materials discovery, super-hard materials, and the inner mineralogy of planets and earth[2, 5, 11, 12]. For instance, CSP has played a pivotal role in identifying the high-temperature superconducting component in hydrogen sulfide under high pressures and unraveling the structures of dense hydrogen[13]. Nonetheless, the considerable time and resource consumption associated with CSP based on density functional theory (DFT) calculations pose limitations on their applicability to a broad array of atomic systems, particularly given the intricacies of conformational space and the expense of local optimization for large systems.

---

[*] Corresponding author. E-mail address: zhaol@scu.edu.cn.
[1] These authors contributed equally to this work.

The landscape of CSP has undergone a transformative shift due to significant strides in recent evolution of machine learning (ML), particularly the emergence of graph neural network (GNN). These innovative approaches have yielded a multitude of methodological breakthroughs and successful applications[14, 15, 16, 17]. ML has been seamlessly integrated into CSP methodologies, demonstrating remarkable acceleration in search efficiency by replacing the expensive DFT calculations[10, 18, 19]. In one such instance, a GNN is harnessed to establish a correlation model between the crystal structure and formation enthalpies ($\Delta H_f$) within a given database[18]. An optimization algorithm is then deployed to expedite the exploration of crystal structure with the lowest formation enthalpy. Furthermore, the filed of CSP has embraced the ML potentials, presenting novel approaches to mitigate the challenges associated with CSP consumption[10, 20]. Recent research showcases the predictive prowess of global optimal search techniques coupled with precise ML potentials in uncovering intricate crystal structures[21]. This approach provides a practical solution to identifying previously elusive dense lithium crystal phases.

The stability metric for CSP is often defined by a complex energy function, potential energy surface, frequently computed using DFT. However, the exchange-correlation functional inherent in DFT methods incorporates multiple approximations that may give rise to notable errors. Additionally, DFT calculations are typically conducted at absolute zero temperature and zero pressure, whereas experiments are performed under finite temperature and pressure conditions. Several studies have illuminated a discernible incongruity between the formation enthalpies predicted by DFT and experimental observations, manifesting discrepancies in the range of 0.08 to 0.18 eV/atom[22, 23, 24, 25]. The disparities between DFT-calculated values and experimental outcomes pose a considerable impediment to accurate ground-state structure prediction as current CSP heavily hinge on DFT-calculated formation enthalpies. In the realm of materials science, where experimental data is often limited, ML models are predominantly trained on DFT-computational datasets. Consequently, ML models inherit the discrepancies between DFT computations and experimental data, further exacerbated by predictive inaccuracies.

While the global energy minimum typically corresponds to experimental realization of a stable crystal structure, nature often favors the selection of specific local-minimum (metastable) structures. Notwithstanding the advantageous formation enthalpies anticipated for numerous materials, their synthesis is often impeded by the intricate interplay of factors, encompassing kinetics and experimental conditions[26]. A notable example is the synthesis of the highly complex and high-energy Hittorf's phosphorus, while other lower energy structures appear to be challenging to synthesize[2, 27]. Consequently, a substantial portion of designed materials faces elimination during the synthesis stage. Conversely, many functional materials with utility are synthesized in metastable states, even when their formation enthalpies do not align with the convex hull minimum[28]. Therefore, integrating synthesizability information of crystal structures into CSP is imperative, serving as a crucial step in streamlining testing costs and accelerating real-world material discovery.

In May 2017, the National Science Foundation sponsored the MGI workshop titled "Advancing and Accelerating Materials Innovation Through the Synergistic Interaction among Computation, Experiment, and Theory: Opening New Frontiers"[29, 30]. Despite the elevated aspirations, the attainment of this objective has only become conceivable in recent times, owing to the emergence of data-driven research methodologies[15, 17]. Most existing methods for CSP overlook the disparity between theoretical predictions and experimental observations. Moreover, these methods predominantly gauge synthesizability based solely on thermodynamic formation enthalpies, disregarding other critical factors influencing synthesis. In response to these limitations, we introduce an efficient CSP method that synergizes optimization algorithms with advanced deep learning models, leveraging a comprehensive DFT materials database and experimental data. We demonstrate the feasibility of predicting material properties closer to true experimental observations using deep transfer learning (TL) models that leverage existing DFT data in conjunction with available experimental data. Going beyond current CSP methods, we develop a model that predicts structures more likely to be synthesized experimentally based on both formation enthalpy and synthesizability information, thus enabling the inverse design of materials. The performance of model has been thoroughly investigated and validated to predict the crystal structures of 17 compounds that have already been experimentally synthesized. The results indicate that incorporating experimental data and synthesizability information significantly improves the accuracy of CSP, reducing disparities with experimental results. This study serves to bridge the long-standing gap between experimental and theoretical realms, potentially opening a new avenue for the inverse design of materials.

**Results**

**Workflow.** A crystal can be identified by three-class descriptors of a material (atomic identity, composition and structure, ACS). In the realm of CSP, the usual objective is to predict the necessary structure based on the given atomic identity and composition. The determination of a crystal's structure, denoted as $\text{Struct}(L, \{R_i\})$, hinges on defining both the lattice constants L and the coordinates $\{R_i\}$ of individual atoms. Although the problem seems clear, uncovering the optimal structure poses a formidable challenge. Optimization algorithms play a pivotal role in mitigating this challenge by enhancing search efficiency, thereby expediting the location of the optimal solution. Incorporating symmetry constraints further refines the search process. The structure with symmetry constraints, denoted as $\text{Struct}_{S,\{W_i\}}(L, \{R_i\})$, benefits from reduced search space through the specification of its space group (S) and associated Wyckoff positions $\{W_i\}$. Simultaneously, we require a ML model capable of swiftly determining the properties of crystal structures with precision, serving as alternatives to traditional DFT calculations.

A specific flowchart is illustrated in Fig. 1a. Initially, based on atomic identities and compositions, we filter a series of conforming space groups and corresponding Wyckoff positions. In the context of CSP optimizations with symmetry constraints, both the space group and corresponding Wyckoff positions become variables, leading to the generation of the S-$\{W_i\}$ two-dimensional discrete space. Under each space group and Wyckoff position, lattice constants and atomic positions become continuous variables, leading to the generation of $\text{Struct}_{S,\{W_i\}}(\{R_i\}, L)$. Subsequently, the crystal structure undergoes conversion into a crystal graph, $G(\{v_i\}, \{e_{ij}\})$, and the ML model is employed to predict the formation enthalpy and the probability of synthesis ($P_{syn}$), producing a fitness value (FV) that reflects the structural quality. In essence, the task of CSP is to identify, through optimization algorithms, the combination of space group, Wyckoff position, lattice constants, and atomic coordinates that minimizes the fitness value. To enhance the accuracy and flexibility of CSP, we conduct a two-stage optimization on these variables. Specifically, we treat continuous variables such as lattice constants and atomic coordinates as one set of optimization parameters, while considering discrete variables such as the space group and Wyckoff position as another set of optimization parameters. During the optimization of lattice constants and atomic coordinates, the objective is to identify the most energetically stable structure. As a result, the fitness value focuses solely on the formation enthalpy. Conversely, for the optimization of the space group and Wyckoff position, the goal is to pinpoint the most likely structure for synthesis among the locally most stable structures. Therefore, the fitness value combines both the formation enthalpy and the probability of synthesis.

**Prediction of formation enthalpy.** For the prediction of formation enthalpies, an effective prediction model should be capable of: (1) distinguishing energetically favorable (low-energy) structures from higher-energy structures; (2) accurately predicting the ground-state structures from a large number of similar structures; (3) exhibiting generalization across a wide range of elements and structures. In this study, a atomistic line graph neural networks (ALGNN) explicitly incorporating the information of bond angles[31] is employed to derive the properties of crystal structures, which are critical for distinguishing many similar structures. See Methods section for details (Fig. 1b-c). The ALGNN model is trained on Open Quantum Materials Database (OQMD v1.5)[23] that includes both ground state and higher-energy structures in a balanced manner, ensuring accurate predictions of formation enthalpies. The training dataset comprises formation enthalpy data for over a million crystal structures, encompassing most elements of the periodic table (see Supplementary Figs. s1-s2). The histogram of the distribution of the formation enthalpies for the entire dataset is shown in Supplementary Fig. s5a.

Initially, we employed the ALGNN model to train formation enthalpies using data from the OQMD dataset. In the test set, the mean absolute error (MAE) and coefficient of determination ($R^2$) of ALGNN model are 0.029 eV/atom and 0.994, respectively, indicating the model's robust performance (Fig. 2a). Notably, the MAE of the current model closely aligns with the 0.033 eV/atom previously reported by Choudhary et al.[31]. Due to the inherent limitations of DFT, as illustrated in Fig. 2b, there is a noticeable discrepancy between the DFT values and the experimental values for some experimentally observed structures (MAE = 0.112 eV/atom). ML models not only inherit the deviation of DFT from experimental values but also introduce additional discrepancies between the model itself and DFT values. Consequently, the gap between the model-predicted formation enthalpy and the experimental value is further accentuated, with an $R^2$ of 0.976 and an MAE of 0.125 eV/atom (Fig. 2c). Despite the challenge posed by the deviation of computations from experiments, direct reliance on experimental data is impractical, given the current difficulty in obtaining substantial amounts of experimental data.

By leveraging deep transfer learning models to learning vast DFT-computational datasets and available experimental values, it becomes feasible to predict material properties with greater proximity to experimental measurements[22]. The ALGNN model is initially training on DFT computational data, typically characterized by high generalization performance owing to its rich data sources. Subsequently, the model undergoes further training using a limited amount of experimental formation enthalpy data. See Methods section for detail (Fig. 1e). Fig. 2d presents a scatter plot of predicted values against true experimental values in the test set. In comparison to the larger deviation observed in the model without experimental data (DFT-ALGNN), the model incorporating experimental data (TL-ALGNN) exhibits a significant enhancement in prediction performance, manifesting as a 27.2% reduction in prediction error. Furthermore, in contrast to the deviation between DFT data and experimental data, the deviation between predicted values form TL-ALGNN model and experimental data is reduced by 18.8%.

**Prediction of synthesis.** Given that the process of material synthesis is intricate and influenced by various factors, the construction of a general theory or a first-principles approach to characterize synthesizability has become impracticable. In this context, we utilize a positive and unlabeled machine learning (PUML) model based on the virtual crystal structures from OQMD and experimentally confirmed crystal from Inorganic Crystal Structure Database (ICSD)[32] to estimate the similarity between candidate crystals and synthesizable crystals, i.e., the probability of synthesis. See Methods section for detail (Fig. 1d). The PUML model, employing the GNN as classifier, has demonstrated significant potential in the analysis of synthesizability for inorganic compounds[33], two-dimensional materials[34] and chalcogenides[35]. The experimentally confirmed crystal structures amount to 37525, encompassing most elements of the periodic table (see Supplementary Figs. s3-s4). The histogram of the distribution of the formation enthalpies for the ICSD data is shown in Supplementary Fig. s5b.

The PUML model involves randomly selecting a portion of the unlabeled data as negative data multiple times (100 in this work) to train the ALGNN classifier. Out of the 1020343 OQMD data points, 37525 unlabeled data points were randomly selected and labeled as negative data to balance with the positive data points (ICSD data). In each iteration, the ALGNN classifier was trained for 50 epochs, achieving convergence in classification performance at a reasonable computational cost (see Supplementary Fig. s6a). Over the course of 100 iterations of training, the average precision on the test set reached 87.1%, a result that aligns well with the outcomes obtained by Jang et al.[33]. This high precision indicates that the model has effectively learned the structural features of synthesizable materials through iterative training. As shown in Supplementary Fig. s6b, a single model exhibits strong prediction precision, and there is a notable increase in classification precision with more classifiers. Consequently, the probability of synthesis for a crystal candidate in CSP is determined as the average of the results from the 100 classifiers.

**Fitness value.** In general, a lower formation enthalpy signifies a more stable crystal structure for a material and, to some extent, may indicate experimental synthesizability. However, the successful experimental synthesis of a crystal is influenced by various factors, such as environmental conditions, and experimental parameters, in addition to thermodynamic stability. Therefore, designing a fitness value that effectively combines both factors of formation enthalpy and probability of synthesis, becomes crucial. The formation enthalpy theoretically ranges from negative infinity to positive infinity, while the probability of synthesis ranges from 0 to 1. To reconcile this disparity, we consider using sigmoid function that maps the formation enthalpy to a bounded range (0 to 1) without sacrificing mathematical monotonicity. Consequently, we designed a fitness value: $FV = \alpha \cdot (1-P_{syn}) + (1-\alpha) \cdot \text{sigmoid}(\Delta H_f)$, where lower FV indicate higher priority. Here, α is a weighting parameter ranging from 0 to 1, allowing for adjustment to provides a flexible approach to balance the influence of formation enthalpy and probability of synthesis in the search for optimal crystal structures.

**Symmetry constraints.** Symmetry plays a pivotal role in defining the properties of a material, influencing its structural morphology, such as structural parameters and atomic arrangement, as well as the physical properties that characterize the material. Nature tends to favor symmetric crystal formations: the prevalence of the asymmetric P1 space group is rare in crystals. The crystal structures exhibit an uneven distribution among space groups[5, 36]. For example, twenty-four out of the 230 potential space groups encompass two-thirds of all inorganic crystals in the ICSD. Characterizing a structure involves two main variables: the lattice constants and the fractional coordinates of all atoms. In this study, we approach CSP under symmetry constraints by introducing two additional structural elements: space group and the Wyckoff position. Specifically, for CSP, one of the 229 space groups with symmetry is randomly selected, and the lattice variables are generated according to the Bravais lattice corresponding to this space group. The atomic positions are determined by Wyckoff's occupation of this space group. These physics-based constraints place candidate structures in more favorable regions of the potential energy surface, significantly simplifying the CSP process. To ensure practical applicability and avoid unreasonable structures, we impose further restrictions on lattice constants, atomic spacing, and cell volume. All interatomic distances must exceed a specified minimum, set at 0.5 times the sum of the radii of the two atoms. Cell angles are constrained to fall between 20° and 160°, and the cell volume is limited to the range of 0.5 to 1.5 times $V_a$, where $V_a$ represents the sum of the volumes of the constituent atoms. These constraints contribute to eliminate redundant and infeasible regions of configuration space, focusing on viable and meaningful regions for

CSP.

**Optimization algorithms.** The parameters of a crystal structure typically involve multiple dimensions, creating a highly complex, high-dimensional parameter space characterized by numerous local minima. Optimization algorithms are instrumental in navigating such intricate spaces, helping to overcome complexity and expedite the discovery of optimal solutions. In this study, we have employed four optimization algorithms: random search (RS), particle swarm optimization (PSO), genetic algorithm (GA), and Bayesian optimization (BO). RS, GA, and PSO have proven successful in DFT-based CSP[3, 5, 7, 10], and BO, which has recently been integrated with ML models for CSP, demonstrates a remarkable ability to identify global minima[18, 37].

Using MgO as an example, we compare the performance of RS, PSO, GA and BO using the DFT-ALGNN-CSP model. Fig. 3 illustrates the evolutionary characteristics of the crystal structure with iteration steps under four different optimization algorithms. The relatively smooth curves over the entire range in the density of states for RS and BO suggest a balance between exploitation and exploration, indicating a high capability to escape certain local minima. Compared to BO, the effective number of attempts in RS is relatively fewer, which may hinder the discovery of the global optimum within a limited number of steps. GA (with mutation probability $P_m$ = 0.1) and PSO (with inertia weight $w$ = 0.8) have a higher number of effective attempts, but most of them are concentrated around local minima, indicating a potential risk of getting stuck in local optima and failing to find the global optimum. The mutation probability of GA and the inertia weight of PSO can control the algorithm's global and local optimization capabilities. Increasing the mutation probability can further improve GA's global search ability but at the expense of reducing the number of effective attempts, as shown in Supplementary Fig s7a. For PSO, reducing the inertia weight worsens the algorithm's global search ability (Supplementary Fig s7b), and a larger inertia weight does not significantly enhance the algorithm's global search ability. In contrast, BO is an algorithm that strikes a balance between exploitation and exploration and has a higher ability to rapidly search for the global optimum within a reasonable time efficiency.

**Validation.** The DFT-ALGNN-BO and TL-ALGNN-BO approach was subsequently employed to identify the experimentally synthesized structures of 17 compounds listed in Table 1. The DFT-ALGNN-BO and TL-ALGNN-BO approach was applied to identify the crystal structures of zinc-blend (ZB)/wurtzite (WZ) and rock-salt (RS), demonstrating the versatility of CSP from ionic to covalent systems. The selected structures also included metallic compounds and semiconductor compounds to showcase the predictive capability of CSP from metals to semiconductors. Furthermore, the chosen structures encompassed elements across seven periods (including the lanthanide and actinide elements), illustrating the generalization ability of CSP across the entire periodic table. Throughout the training, validation, and testing processes, the data of the 17 compounds studied in this work were excluded. BO was employed for the optimization of space group and Wyckoff positions within 200 steps, while lattice constants and atomic coordinates were optimized within 300 steps. We considered CSP successful if the top five crystal structures in the ranking of fitness value (lower fitness value indicating higher priority) contained the target structure, i.e., the experimentally validated structure.

From Table 1, several observations can be made: (1) The success rate of fixed crystal system searches is significantly higher than that of all crystal system searches, regardless of the value of α. (2) Whether in fixed crystal system searches or all crystal system searches, α=0.5 (i.e., considering both energy and synthesizability assessments) yields a higher success rate. (3) Integrating energy and synthesizability assessments in fixed crystal system searches can achieve a relatively high success rate (82.4%). The successful applications of ALGNN-BO in cubic crystal systems (HLi), hexagonal crystal systems (SrCu), tetragonal crystal systems (AuCl), and ternary compounds ($BaZrO_3$) demonstrate the applicability of ALGNN-BO in various crystal systems and multi-component compounds. Among the numerous compounds, HLi and $ThO_2$ present interesting cases. Taking $ThO_2$ as an example, as shown in Supplementary Fig. s8, DFT-ALGNN-BO with α=0 found the ground-state structure in 62 steps, which, however, was not experimentally validated. When further considering the synthesizability of the crystal (α=0.5), the optimal structure predicted by TF-ALGNN-BO aligns with the experimentally confirmed structure. However, assigning more weight to crystal synthesizability does not necessarily mean that the identified structure is more likely to be experimentally synthesized. For instance, RbI and LaTe can only be correctly identified when considering only the formation enthalpy, indicating a potential misjudgment in the PUML model (Precision=0.87).

Fig. 4 depicts the percentage errors of formation enthalpy and lattice constants predicted by TL-ALGNN-BO, DFT-ALGNN-BO, and those obtained from the OQMD database compared to experimental values for 15 compounds. The prediction results of AuCl and CuSr can be found in Supplementary Fig. s9 (the predictive results are relatively subpar). For the 15 compounds, the lattice constants from TL-ALGNN-BO, DFT-ALGNN-BO, and the OQMD database fall within a 10% error range relative to experimental values, as shown in Fig. 4b. Specifically, the lattice constants from DFT calculations align quite well with experimental values, with an average error of less than 2% (1.83%). The lattice constants predicted by TL-ALGNN-BO, which incorporates some experimental data, have an average error of 2.02%, slightly higher than DFT computations but significantly improved compared to DFT-ALGNN-BO model (2.51%). In comparison to lattice constants, both ML-CSP model predictions and DFT calculations in the OQMD database exhibit significant discrepancies in formation enthalpy relative to experimental values. Specifically, the average error between DFT and experimental data in formation enthalpy is 10.74%, while DFT-ALGNN-BO increase this difference to 11.28%. Using TL-ALGNN-BO for prediction reduces the error to 8.34%, consistent with the analysis in Fig. 2.

**Discussion**

Currently, the determination of the microscopic structure of materials is primarily achieved through various experimental techniques, such as X-ray diffraction and nuclear magnetic resonance. However, experimental measurements are often subject to various limitations, such as the purity of the experimental sample and the strength of the experimental spectrum. As a result, relying solely on experimental methods to determine the structure of materials presents some challenges. On the other hand, one might wish to ascertain whether, for a given stoichiometry, a given experimental formation enthalpy precisely corresponds to a crystal structure reported in the ICSD. However, this comparison is not feasible because thermodynamic databases often lack explicit crystal structure information. Therefore, the development of theoretical CSP methods has been a long-standing expectation in the fields of physics, chemistry, and materials research. Traditional DFT-CSP and ML-CSP typically rely on theoretically calculated formation enthalpies and utilize optimization algorithms to search for the ground-state structure with the lowest energy. However, current CSP methods do not take into account the differences between experimentally obtained formation enthalpies and those from theoretical calculations. Moreover, the synthetic process of materials is inherently intricate, influenced not only by thermodynamic stability but also by a myriad of factors such as kinetics, synthesis techniques, and more. Our model utilizes experimental formation enthalpy data to significantly reduce the differences between theoretical and experimental values. Additionally, it employs an analysis of structural synthesizability to ensure that the predicted structures are highly likely to be experimentally validated. To the best of our knowledge, this study represents the first attempt to establish a DFT-Experiment-ML framework for CSP, bridging the gap between theory and experiment.

In comparison to the OQMD v1.3 and Materials Project[38] used by Cheng et al., we utilize the OQMD v1.5 dataset, which contains information on over one million structures. In contrast to CGCNN[39] developed by Xie et al. and MEGNet[40] used by Cheng et al., the ALGNN used in this work further considers the information of crystal bond angles. Since the expansion of training data and improvements to the prediction model, the current ALGNN-CSP demonstrates strong generalization across most elements in the periodic table, while significantly enhancing prediction accuracy. For example, CGCNN is not able to discern some very similar structures (see Supplementary Fig. s10); however, ALGNN demonstrates the capability to identify them. Previous research has indicated that the BO algorithm, when combined with CSP, outperforms PSO and RS, which are often combined with DFT-CSP[18]. We further conducted an in-depth comparison of BO, PSO, RS, and GA algorithms in CSP. The results demonstrate that BO is an algorithm that strikes a balance between exploitation and exploration, and it has a higher ability to rapidly search for the global optimum within a reasonable time efficiency.

Termed as the forward "properties follow structures" approach, the existing methodology for material design harnesses the structural knowledge of a material as input to forecast the ensuing material characteristics. However, this trial-and-error process is both time-consuming and labor-intensive. In response to these limitations, the inverse design paradigm (structures follow properties) has emerged, aiming to generate candidate structures that correspond to desired material properties[41, 42, 43, 44, 45, 46]. In this study, we coupled the synthesizability information of crystals in CSP, enabling the inverse design of crystal structures that are more likely to be experimentally synthesized. This approach significantly accelerates material design and reduces the cost of experimental validation. In fact, our model can achieve more than CSP alone, as it can be applied to the general inverse design of materials by incorporating additional target functions as conditions. For instance, one can add information such as thermal conductivity or optoelectrical conductivity in CSP to expedite the discovery of new materials with enhanced thermal conductivity or optical-electrical properties.

The application of ML methods in the field of CSP is still in its early stages of development, with numerous scientific and technical challenges awaiting resolution. A prevalent issue in CSP is the accurate prediction of complex systems, a challenge that has persisted for a long time. Our results also indicate that the current ALGNN-CSP has significant room for improvement in predicting complex systems. On the other hand, compared to the cubic crystal system, the predictive capability and performance of ALGNN-CSP in other non-cubic crystal systems are significantly reduced. This can be mainly attributed to the relatively low proportion of other non-cubic crystal systems in the training data, as indicated in supplementary Fig. s11a. However, in nature, specifically in the crystal structures confirmed in the ICSD, other non-cubic crystal systems constitute a considerable proportion (see supplementary Fig. s11b). An apparent avenue for improvement is to enhance the overall predictive accuracy of CSP by improving the predictive precision of the model. However, apart from refining neural network models, a more meaningful improvement could involve further supplementing and refining the database, particularly with additional structural data for other non-cubic crystal systems. Similarly, additional experimental data will further diminish the disparities between CSP results and experimental outcomes.

In summary, we introduce an efficient CSP method that combines optimization algorithms with advanced deep learning models, leveraging extensive DFT material databases and experimental data. We utilize an ALGNN model that incorporates the information of bond angle, enabling accurate mapping between structure and properties. Through deep transfer learning techniques to learn experimental formation enthalpy, our model can predict the formation enthalpy of structures with greater accuracy, aligning more closely with experimental values. By integrating formation enthalpy with synthesizability, the ALGNN-CSP model can predict the crystal structures that not only exhibit thermodynamic stability but are also better suited for practical synthesis under real-world laboratory conditions. Simultaneously, we evaluate the performance of different optimization algorithms in CSP, confirming that BO can assist the model in more efficiently exploring potential structures within the vast chemical space. The model is subsequently applied to predict the crystal structures of 17 typical compounds, showcasing its ability to generalize across a spectrum of compounds, spanning from ionic to covalent, metals to semiconductors, and encompassing the entirety of the periodic table. The predicted results show that the CSP model, incorporating experimental formation enthalpy data, significantly reduces the disparities between predicted and experimental values of formation enthalpy and lattice constants. Compared to CSP based solely on formation enthalpy, the addition of synthesizability analysis greatly improves the model's ability to identify crystals that can be experimentally synthesized. This study bridges the long-standing gap between experimental and theoretical realms, potentially opening a new avenue for the inverse design of materials.

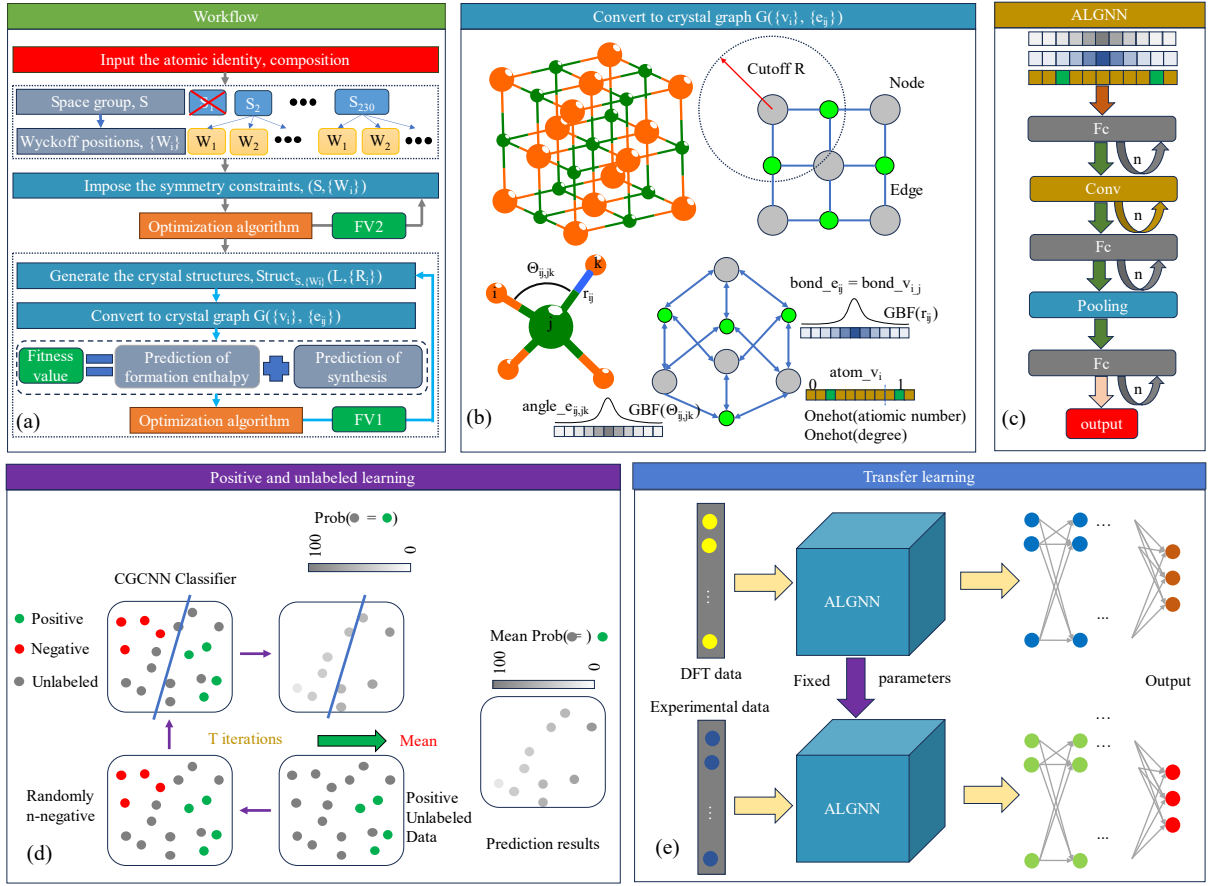

**Fig. 1. Flowchart and sub-modules of CSP.** The flowchart of the CSP with optimization algorithm approach is in **(a)**. The flowchart comprises four main modules. **(b)** crystal graph generation: involving the generation of crystal graphs and representing the structural information of materials; **(c)** Architecture of ALGNN: describing the framework of the AlGNN used for modeling and learning complex relationships between crystal structures and material properties; **(d)** PUML model: referring to a learning paradigm that involves the training of the model using positive instances and unlabeled instances, enabling the model to identify potentially favorable structure. **(e)** Transfer learning: by harnessing vast DFT-computational datasets and available experimental observations, it becomes feasible to predict material properties with greater proximity to genuine experimental measurements.

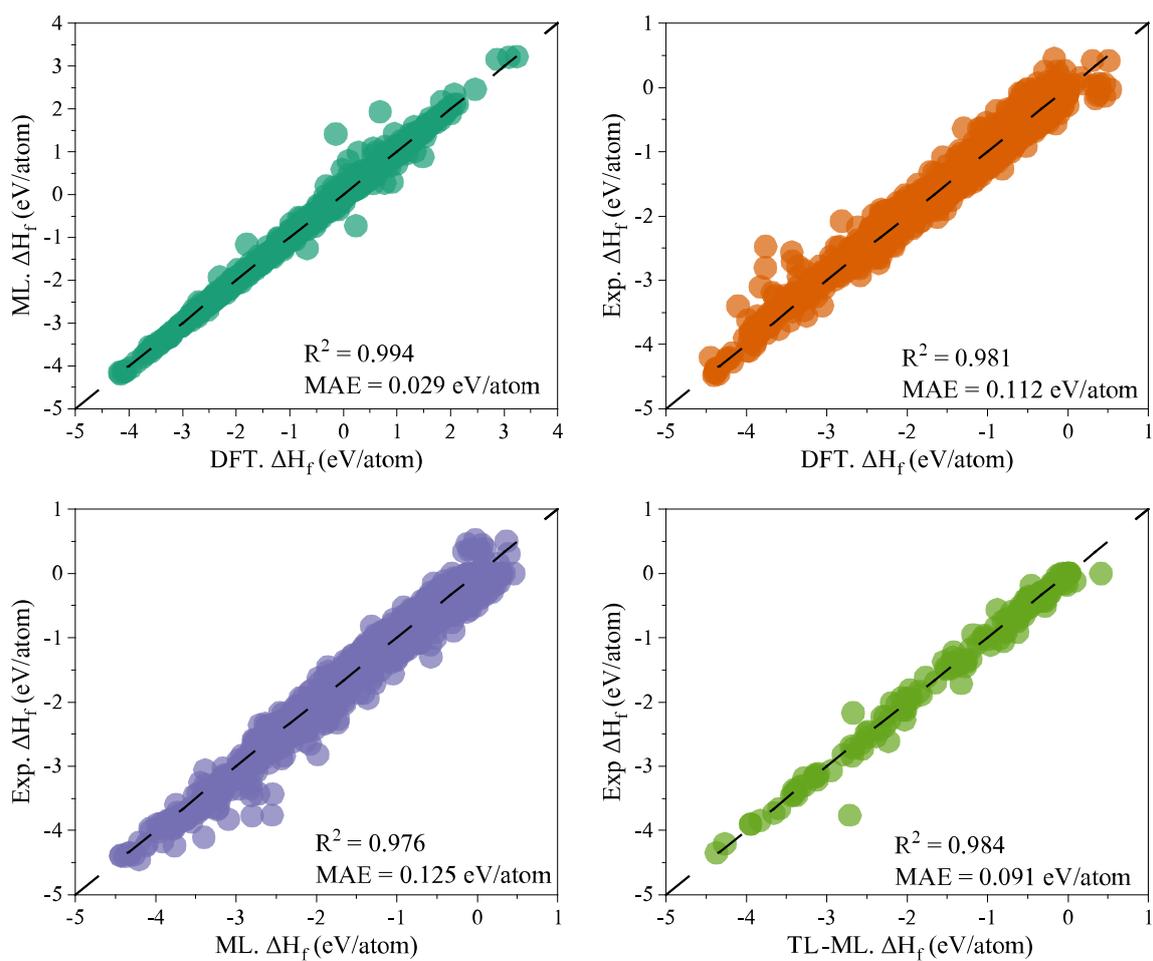

**Fig. 2. The performance of ALGNN model. (a)** Comparison between DFT-ALGNN-predicted and DFT-calculated formation enthalpies. **(b)** Comparison between DFT-calculated and experimental formation enthalpies. **(c)** Comparison between DFT-ALGNN-predicted and experimental formation enthalpies. **(d)** Comparison between TL-ALGNN-predicted and experimental formation enthalpies.

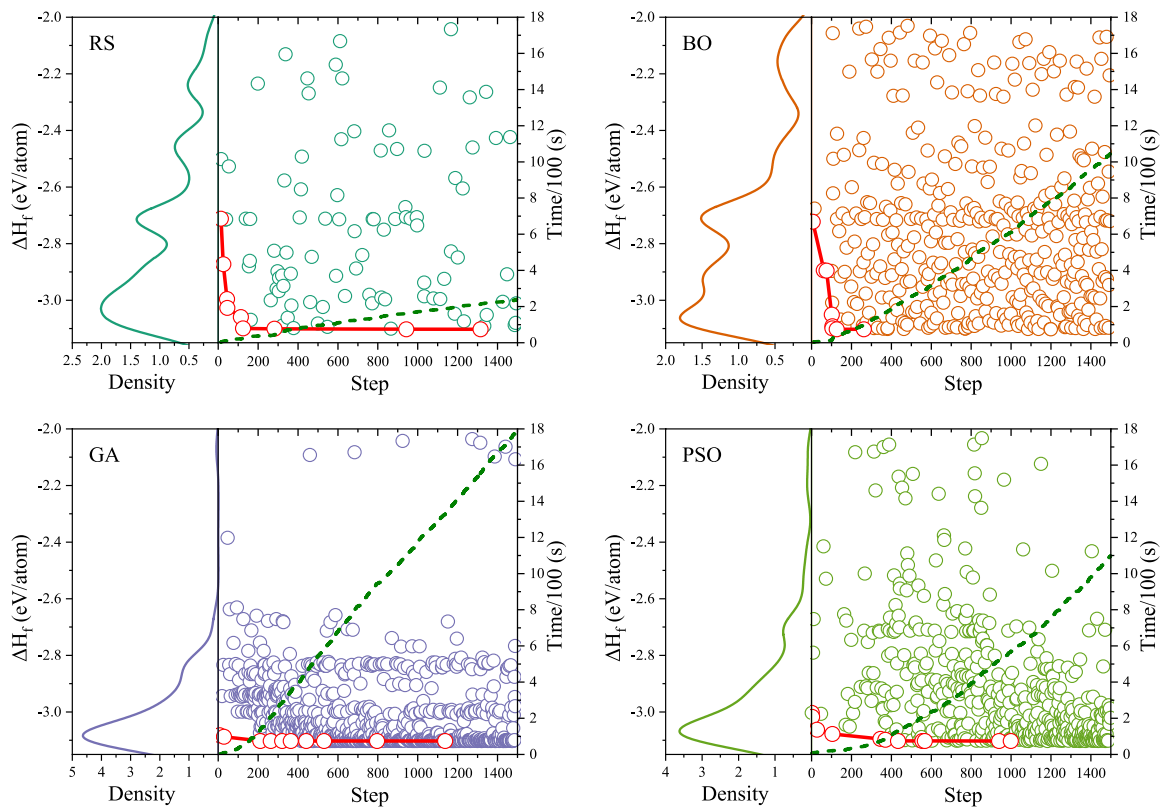

**Fig. 3. The process and performance of DFT-ALGNN-CSP for four optimization algorithms.** The green dashed line represents the cumulative time required by each optimization algorithm, while the red solid line depicts the variation of the most stable structure over time. The left panel displays the density of states at the energy level, providing insights into the distribution of structures at different energy levels during the optimization process.

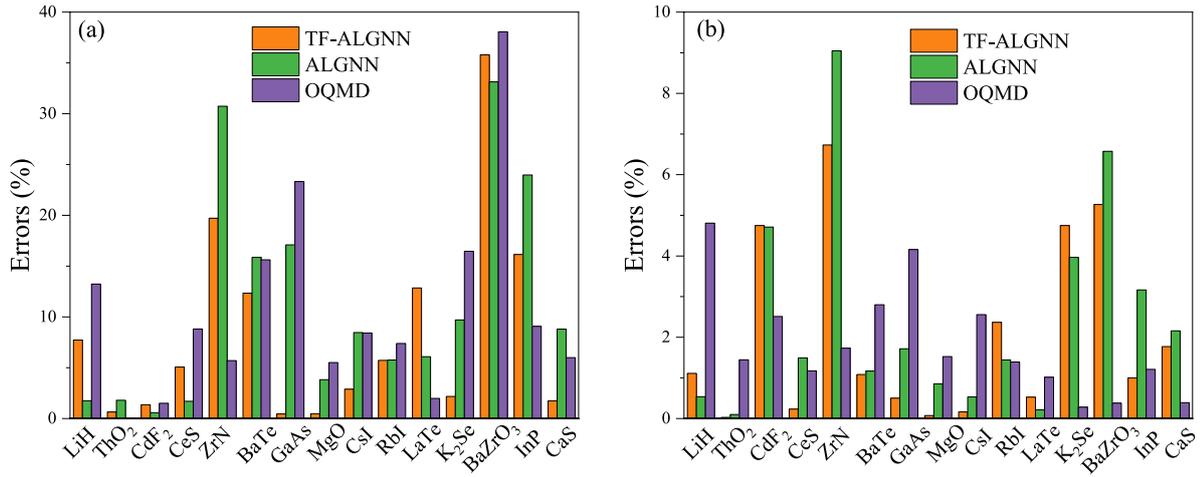

**Fig. 4. The comparisons of structures derived from ML model predictions and QOMD with experimental data.** The discrepancies in formation enthalpy **(a)** and lattice constants **(b)** predicted by TL-ALGNN-BO, DFT-ALGNN-BO, and those obtained from the OQMD database compared to experimental values for 15 compounds.

| Compositions | All crystal system | | | Fixed crystal system | | |
| --- | --- | --- | --- | --- | --- | --- |
| | $\alpha = 0$ | $\alpha = 0.5$ | $\alpha = 1$ | $\alpha = 0$ | $\alpha = 0.5$ | $\alpha = 1$ |
| LiH | | √ | √ | | √ | √ |
| $ThO_2$ | | | | | √ | √ |
| $CdF_2$ | √ | √ | | √ | √ | |
| CeS | √ | √ | √ | √ | √ | √ |
| ZrN | √ | √ | √ | √ | √ | √ |
| BaTe | | √ | √ | | √ | √ |
| GaAs | √ | √ | √ | | √ | √ |
| MgO | √ | √ | | √ | √ | √ |
| AuCl | | | | √ | | |
| CsI | √ | √ | √ | √ | √ | √ |
| RbI | √ | | | √ | | |
| LaTe | √ | | | √ | | |
| $K_2Se$ | | | | √ | √ | |
| $BaZrO_3$ | | | | √ | √ | √ |
| InP | | √ | √ | | √ | √ |
| SrCu | | | | √ | √ | |
| CaS | √ | √ | √ | √ | √ | √ |
| Accuracy | 52.9% (9/17) | 58.8% (10/17) | 47.1% (8/17) | 70.6% (12/17) | 82.4% (14/17) | 64.7 % (11/17) |

Table 1 The performance of ALGNN-BO with different search strategies (All crystal system or Fixed crystal system) and fitness values ($\alpha$ = 0, 0.5, and 1) for CSP of 17 typical compounds.